\documentclass[twocolumn,preprintnumbers,amsmath,amssymb]{revtex4}
\usepackage{graphics}
\usepackage{epsfig}
\usepackage{epstopdf}
\usepackage{color}
\usepackage{subfigure}
\usepackage{placeins}

\begin{document}

\title{Direct Spectro-Temporal Characterization of Femtosecond Extreme-Ultraviolet Pulses}
\author{David Gauthier$^{1}$, Beno\^{i}t Mahieu$^{1,2,3}$, Giovanni De Ninno$^{1,2}$}
\affiliation{
1. Laboratory of Quantum Optics, Nova Gorica University, Nova Gorica, Slovenia\\
2. Sincrotrone Trieste, Trieste, Italy \\
3. Service des Photons Atomes et Mol\'{e}cules, Commissariat \`{a} l'Energie Atomique, Centre d'Etudes de Saclay, Gif-sur-Yvette, France\\
}

\date{\today}

\begin{abstract}
We propose a method for a straightforward characterization of the temporal shape of femtosecond pulses in the extreme-ultraviolet/soft X-ray spectral region. The approach is based on the presence of a significant linear frequency chirp in the pulse. This allows to establish an homothetic relation between the pulse spectrum and its temporal profile. The theoretical approach is reminiscent of the one employed by Fraunhofer for describing far-field diffraction. As an application, we consider the case of a seeded free-electron laser (FEL). Theory is successfully benchmarked with numerical simulations and with experimental data collected on the FERMI@Elettra FEL. The proposed method provides FEL users with an on-line, shot-to-shot spectro-temporal diagnostic for time-resolved experiments.
\end{abstract}

\maketitle

New light sources generating extreme-ultraviolet/soft X-ray (XUV) femtosecond pulses are among the most powerful tools for investigating the fundamental properties of matter, in gas, liquid or solid-state phases \cite{neutze,young,meyer}. In fact, short wavelength (i.e., high-energy) radiation allows to achieve atomic spatial resolutions and to probe core levels of atomic and molecular bound states. Short pulses are instead required for the study of ultra-fast chemical and physical reactions, allowing separate access to electron, spin and lattice relaxation constants.
\\
A large number of these studies rely on pump-probe techniques \cite{zewail}, in which two light pulses with adjustable time delay and different wavelengths are used to investigate the dynamics of a given process. For the success of such experiments, a detailed knowledge and a complete control of the spectro-temporal properties of light pulses are essential. This should include the possibility for users to monitor and control the spectro-temporal features of XUV femtosecond pulses in \emph{real time} (i.e., during experiments).
\\
Nowadays, measuring the spectrum of XUV pulses is not an issue, whereas getting access to the temporal information is a more challenging task. Indeed, several methods have been proposed to measure the phase and shape of femtosecond pulses in the infrared, visible and near ultraviolet spectral ranges \cite{trebino,iaconis}. However, in the XUV domain, the strong absorption of solid-state crystals does not enable non-linear optical effects on which temporal characterizations usually rely, making them significantly more difficult. Usually, the duration of XUV pulses is determined by means of cross-correlation schemes, based on the photo-ionization of an atomic gas by the measured XUV pulse, \textquotedblleft dressed\textquotedblright by a time-delayed femtosecond infrared laser \cite{schins,glover,toma}. Based on this and other approaches, several methods have been developed for the temporal characterization of ultra-short pulses generated by single-pass free-electron lasers (FEL) \cite{radcliffe,drescher2,grguras} and high harmonic generation (HHG) sources \cite{constant,drescher1,sekikawa2,mairesse}. However, up to now, the measurement of the temporal shape of femtosecond pulses in the XUV spectral range is not a trivial issue and requires a complex and normally dedicated experiment.
\\

In this Letter we present a method allowing a fast on-line characterization of the temporal shape of XUV femtosecond pulses, generated by a single-pass FEL. In the considered configuration, the FEL is seeded by a laser pulse, carrying a controllable frequency chirp. The method goes as follows. We first demonstrate that, under specific constraints, the presence of a significant linear frequency chirp in an optical pulse allows to establish a direct homothetic relation between its spectral and temporal profiles. Then, we study the case of a seeded FEL and we demonstrate that, under the specified conditions, the duration and the temporal shape of the FEL pulse can be simply retrieved on-line by a measurement of its spectrum. The conditions of validity of the proposed technique are determined by means of a rigorous analytical method, reminiscent of the one employed by Fraunhofer to define the diffraction pattern of a monochromatic wave in paraxial far-field approximation \cite{goodman}. The analytical result is validated first by means of numerical simulations and then by a direct comparison with experimental data collected on the FERMI@Elettra FEL \cite{allaria}.
\\

\begin{figure}[!h]
\centering
\includegraphics[height=60mm]{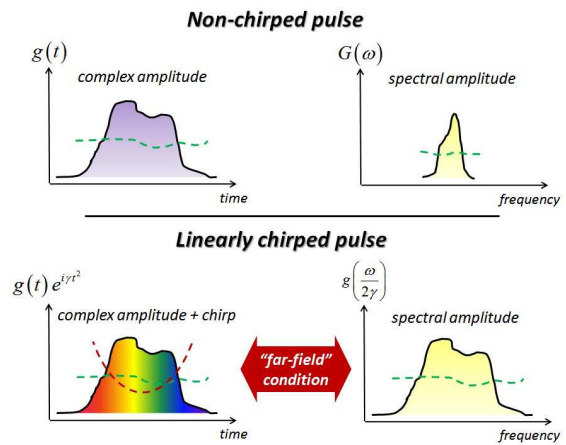}
\caption{Schematic representation of the \textquotedblleft far-field\textquotedblright condition which allows to find a correspondence between the temporal and the spectral profiles of a linearly chirped optical pulse. The continuous lines represent the envelopes and the dashed lines represent the phases.
\label{fig1}}
\end{figure}

In order to illustrate the method, we start by considering a pulse characterized by the electromagnetic field $f(t)=g(t)e^{i\gamma t^2}e^{i\omega _0 t}$, where $g(t)$ is a generic complex amplitude, $\gamma t^2$ is a known (controllable) quadratic phase term and $\omega_0$ is the pulse central frequency. The quadratic term induces a temporal dispersion of the pulse spectral component, according to the following relation:
\begin{equation}
\omega_{ins}(t)=2\gamma t+\omega_0
\label{omegainst}.
\end{equation}
Here $\omega_{ins}(t)$ represents the instantaneous angular frequency along the pulse at a given temporal position $t$. Such a frequency is the result of the interference of multiple spectral components. The number of spectral components contributing to $\omega _{ins}(t)$ is large for small temporal dispersion (i.e., small $\gamma$) and decreases for increasing temporal dispersion (i.e., large $\gamma$). For sufficiently large $\gamma$ values, one can approximately associate a single spectral component to each temporal position in the pulse. When this occurs, the temporal form of the pulse becomes similar to its spectral shape.
\\
In order to provide a quantitative description of this phenomenon, we exploit a parallelism between the problem in hand and the well-known problem of spatial-frequency dispersion in diffraction. For our demonstration, we will make use of hypotheses similar to the ones employed by Fraunhofer for describing the far-field diffraction \cite{goodman}.
\\
The pulse spectrum is given by:
\begin{equation}
X(\omega )\simeq \left|\mathcal{F}_{\omega }\left\{g(t)e^{i\gamma t^2}\right\}\right|^2
\label{spectre1},
\end{equation}
where $\mathcal{F}_{\omega }\{f(t)\}=\underset{-\infty }{\overset{+\infty }{\int }}dt\, f(t)e^{-i\omega t}$ describes the Fourier transform of the function $f(t)$, from the temporal variable $t$ to the frequency variable $\omega$ \footnote{for simplicity, in Eq. (\ref{spectre1}) the constants are neglected and the angular frequency variable $\omega$ is centred around the central frequency $\omega_0$.}. Using Fourier formalism \cite{goodman}, we can write $X(\omega)$ as a convolution product:
\begin{equation}
X(\omega)=\left|G(\omega)*e^{-i\frac{\omega^2}{4\gamma}}\right|^2
\label{spectre2},
\end{equation}
where $G(\omega)$ represents the Fourier transform of the complex amplitude $\mathcal{F}_{\omega }\left\{g(t)\right\}$ and $e^{-i\frac{\omega^2}{4\gamma}}$ is the Fourier transform of the quadratic phase term $e^{i\gamma t^2}$. At this step, an analogy can be drawn between the spectrum $X(\omega)$ and the one-dimensional diffraction pattern originated from a \textquotedblleft transversal field distribution\textquotedblright $G(\omega)$. In the previous relation, the convolution with the exponential term $e^{-i\frac{\omega^2}{4\gamma}}$ gives rise to a phenomenon similar to the longitudinal space propagation of a diffracted transverse field distribution. In paraxial approximation, such a propagation generates a linear dispersion of the spatial-frequencies components. In our case, this corresponds to the linear frequency dispersion in the pulse, caused by the quadratic phase term $\gamma t^2$. After the development of the convolution product, $X(\omega)$ can be written as an inverse Fourier transform, $\mathcal{F}^{-1}$, in the variable $\frac{\omega}{2\gamma}$:
\begin{equation}
X(\omega )=\left|\mathcal{F}_{\frac{\omega }{2\gamma }}{}^{-1}\left\{G(\omega ')e^{-i\frac{\omega '^2}{4\gamma }}\right\}\right|^2
\label{spectre3}.
\end{equation}
Let us focus on the exponential term $e^{-i\frac{\omega'^2}{4\gamma}}$. In far-field diffraction, according to Fraunhofer, this phase term varies slowly in the Fourier integration domain where the square integral function $G(\omega)$ is different from zero. This condition is equivalent to:
\begin{equation}
\frac{\sigma_{G}^2}{4\gamma}=N\ll 1
\label{condition N},
\end{equation}
where $\sigma_{G}$ represents the standard deviation of the function $G(\omega)$ and $N$ is the Fresnel number \cite{goodman}. When the condition Eq.~(\ref{condition N}) is fulfilled, one can reasonably approximate the exponential term with unity. As a result, Fraunhofer diffraction pattern provides a direct representation of the Fourier transform of the initial field distribution. In our case, since $G(\omega)$ is the Fourier transform of the complex amplitude $g(t)$, we finally get:
\begin{equation}
X(\omega)=\left|g\left(\frac{\omega}{2\gamma}\right)\right|^2
\label{homothetie}.
\end{equation}
We thus obtain the following noticeable result: under the \textquotedblleft far-field\textquotedblright condition Eq.~(\ref{condition N}), the spectrum of a linearly chirped optical pulse provides a direct representation of its temporal shape, through the variable transformation $t=\frac{\omega}{2\gamma }$. A schematic representation of the proposed spectro-temporal equivalence is given in Fig.~\ref{fig1}.
\begin{figure*}[!ht]
\centering
\includegraphics[height=40mm]{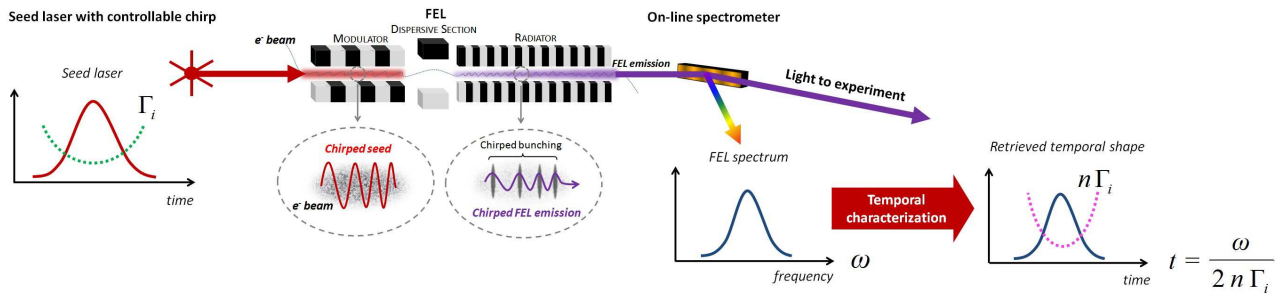}
\FloatBarrier 
\caption{Schematic layout of a seeded FEL and sketch of the proposed method to obtain a real-time spectro-temporal characterization.
\label{fig2}}
\end{figure*}
\\

Let us now apply the previous method to determine the temporal profile of light pulses generated by a single-pass seeded FEL working in the coherent harmonic generation regime \cite{yu1,yu2}. In a seeded FEL (see Fig.~\ref{fig2}), a relativistic electron beam propagating trough an undulator (called modulator) interacts with a collinear externally-injected optical pulse (the seed). The interaction modulates the electron-beam energy. Energy modulation is then transformed into spatial bunching, when the electron beam propagates through a magnetic chicane (dispersive section). The bunching (as the energy modulation) has a periodicity determined by the seed frequency $\omega_0$. In addition, it also presents significant components at the harmonics of the latter, i.e., at $n\omega_0$ (where $n$ is an integer). Finally, the bunched electron beam is injected into a long undulator chain (called a radiator), where it emits coherently at one of the seed harmonics. In the radiator, the electromagnetic intensity generated by the bunched electrons is amplified, until when, due to bunching deterioration, electrons are no longer able to supply energy to the wave and the process reaches saturation.
\\
The seed is normally a Gaussian quasi-monochromatic pulse (e.g., at low-order harmonic of a Ti:Sa laser), with temporal width $\sigma _{Seed}$. By propagating the seed through a stretcher (prior to injection into the modulator), one can induce and control a linear frequency chirp, defined by the coefficient $\Gamma _i$ in the following relation. The resulting optical pulse reads:
\begin{equation}
Seed(t)=e^{-\frac{t^2}{2\sigma _{Seed}{}^2}}e^{i\Gamma _it^2}e^{i\omega _0t}
\label{seed}.
\end{equation}
In this case, the periodic modulation of the bunching is determined by the instantaneous frequency $\omega_{ins}(t)$ of the seed laser, according to Eq.~(\ref{omegainst}). In turn, the FEL emission at the harmonic $n$ will be chirped at the instantaneous frequency $n\omega_{ins}(t)$. As a result, the quadratic phase of the seed laser (multiplied by the harmonic order $n$) is \textquotedblleft transferred\textquotedblright to the FEL pulse \cite{stupakov,geloni,ratner}. The latter can therefore be written as
\begin{equation}
FEL(t)=g(t)e^{in\Gamma_it^2}e^{in\omega _0t}
\label{FEL},
\end{equation}
where $g(t)$ represents the FEL complex amplitude. This expression is analogous to the one considered previously to represent a generic linearly chirped pulse; here $\gamma = n\Gamma_i$.
\\
The complex amplitude $g(t)$ contains a distorted phase, adding to the controlled term $n\Gamma_i t^2$. Most critical distortions of phase are mainly due to the presence of a non-linear curvature in the electron-beam energy profile and may take place during the bunching construction and/or as a result of the amplification process in the radiator \cite{stupakov,lutman}. Additional (less critical) possible sources of distortion are the presence of high-order non-linear terms in the phase of the seed and the intrinsic FEL phase \cite{murphy,wu}. Another important distortion of the complex amplitude can appears if the FEL is run in over-saturated regime related to the bunching destruction. Distortions may in principle increase the initial spectrum width $\sigma_G$ and, therefore, spoil the \textquotedblleft far-field\textquotedblright condition (Eq.~(\ref{conditionFEL})). However, as it will be clear in the following, it is reasonable to assume that, under the conditions of good seed quality, smooth electron-beam energy distribution, slightly saturated FEL regime and large $\Gamma_i$, this effect can be neglected.
\\
The possibility of controlling $\Gamma_i$ allows to perform a complete on-line characterization of the FEL pulse. In slightly saturated regime, one can assume the FEL pulse to have a Gaussian profile with duration $n^{\frac{-1}{3}}\sigma_{Seed}$ \cite{stupakov,ratner}. In this case, if we neglect the distorted phase of the complex amplitude, the width of $G(\omega)$ is given by $\sigma_G=\frac{1}{n^{\frac{-1}{3}}\sigma_{Seed}}$, and we can re-write relation (\ref{condition N}) only in terms of the seed laser properties:
\begin{equation}
\frac{\sigma_G{}^2}{4n\Gamma_i}=\frac{1}{4n^{\frac{1}{3}}\sigma_{Seed}{}^2\Gamma_i}=N\ll 1
\label{conditionFEL}.
\end{equation}
This relation offers a criterion to ensure the validity of the proposed approach in function of the seed parameters as well as of the harmonic $n$ of the FEL emission.
\begin{figure}[!h]
    \centering
    \subfigure
    {
        \includegraphics[clip=true, height=30mm]{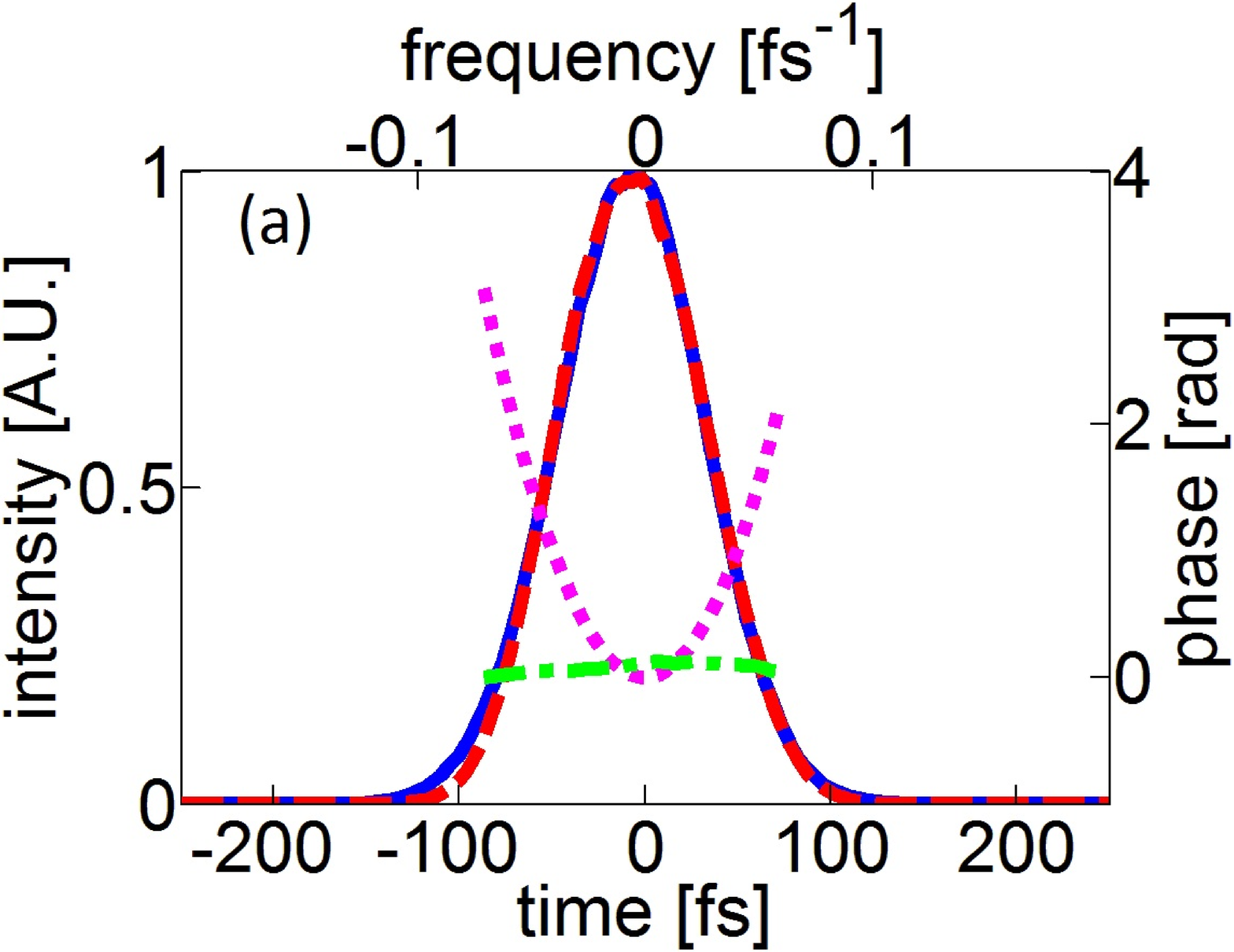}
        \label{figure3a}
    }
    \subfigure
    {
        \includegraphics[clip=true, height=30mm]{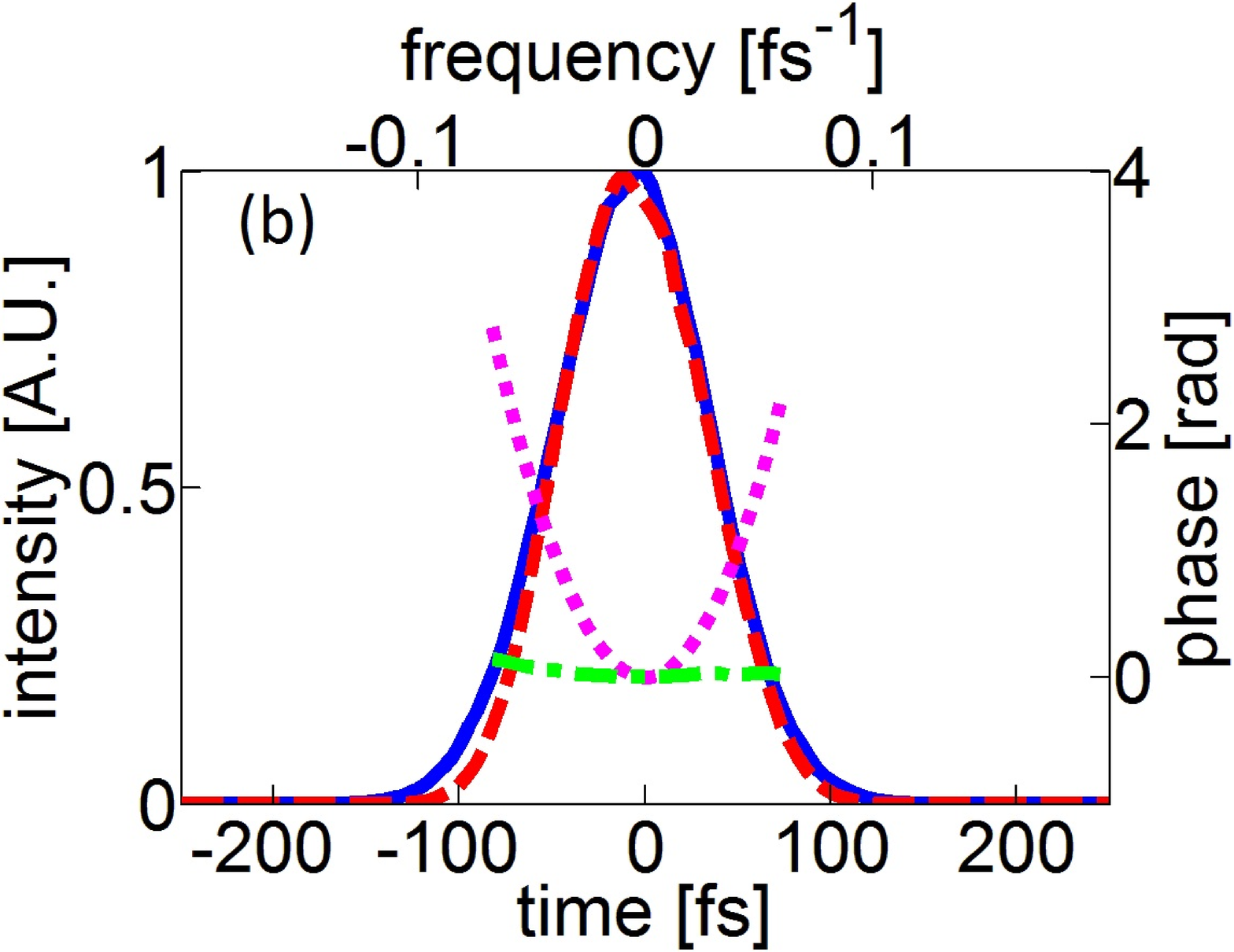}
        \label{figure3b}
    }
    \subfigure
    {
        \includegraphics[clip=true, height=30mm]{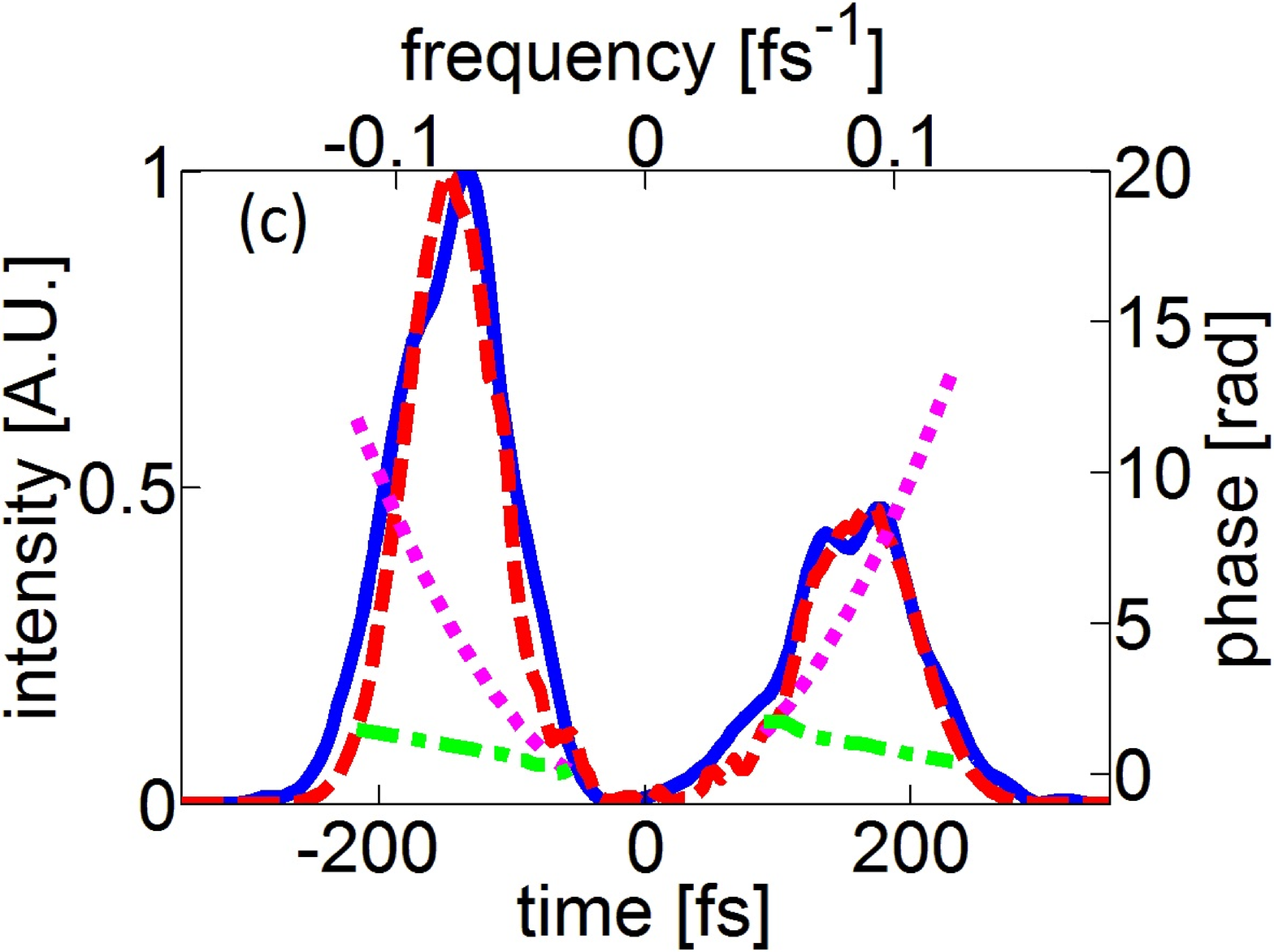}
        \label{figure3c}
    }
    \subfigure
    {
        \includegraphics[clip=true, height=30mm]{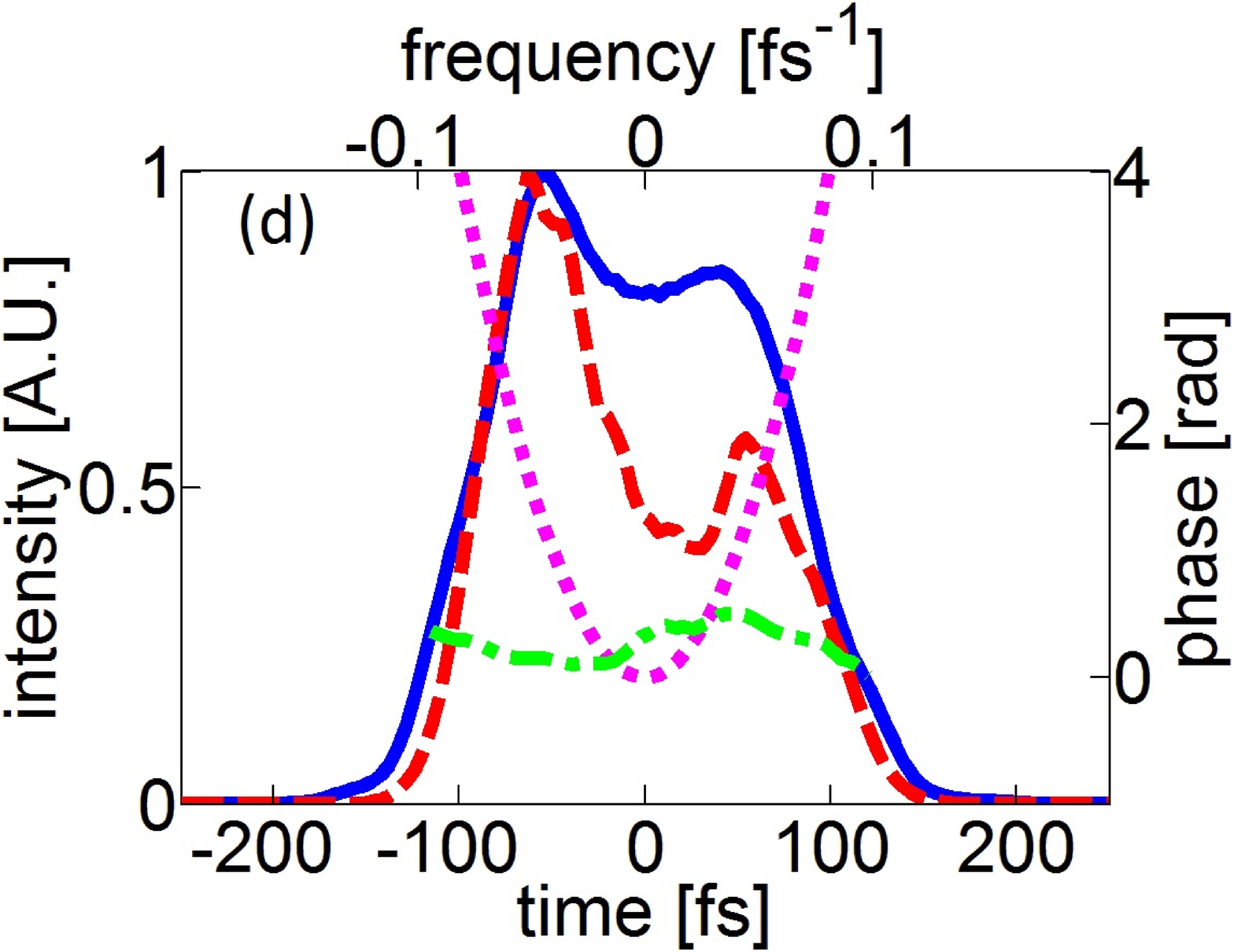}
        \label{figure3d}
    }
\FloatBarrier 
\caption{Comparison between the temporal (dashed red line) and spectral (solid blue line) profiles  obtained using the numerical code PERSEO: (a) slightly saturated regime with ideal electron beam; (b) slightly saturated regime with quadratic electron-beam energy profile (energy curvature: $5\,\text{MeV/ps}^2$); (c) double-pulse regime; (d) over-saturated regime; The temporal scale (bottom axis) is obtained from the spectral one (top axis), using the homothetic transformation given in Eq.~(\ref{homothetie}). The dotted curves represent the quadratic phase term $n\Gamma_i t^2$ and the dot-dashed curves correspond to the other phase terms of the pulse.
\label{fig3}}
\end{figure}
\\

In order to check our results, we performed a campaign of numerical simulations using, as a reference, the case of the FERMI@Elettra FEL \cite{allaria}. In this case, the seed signal, provided by the third harmonic of a Ti:sapphire laser \cite{danailov}, is centered at 261 nm, it has a Gaussian shape and a bandwidth of about 0.9 nm (FWHM). By modulating the seed spectral phase, one can decide the value of $\Gamma_i$ and the correspondent pulse duration, $\sigma _{Seed}$.
\\
Figure~\ref{fig3} reports the spectral and temporal FEL output pulses, obtained using the numerical one-dimensional code PERSEO \cite{perseo} \footnote{Simulations performed using three-dimensional codes like, e.g., GENESIS [S. Reiche, {\em Nucl. Inst. and Meth. A\/} {\bf 429}, 243 (1999)], provide similar results.}, at the harmonic order $n=8$, for different operation regimes. For the case of Fig.~\ref{figure3a} the seed power and the strength of the dispersive section have been tuned so as to reach saturation at the end of the radiator chain. This is the standard configuration in which a seeded FEL is normally operated. For the case of FERMI@Elettra, the standard seed laser parameters ($\sigma _{Seed}=122\:\text{fs}$, $\Gamma_i=5.1 \:10^{-5}\:\text{fs}^{-2}$) correspond, according to Eq.~(\ref{conditionFEL}), to $N=0.16$. As it can be seen, our prediction (Eq.~(\ref{homothetie})) is fully confirmed: the spectral (continuous) and temporal (dashed) profiles are very similar. Note that the induced quadratic phase (dotted curve) dominates the phase distortion (dash-dot line). Simulations confirm that the agreement becomes worse for increasing values of the parameter $N$ (data not shown).
\\
As it was mentioned, non-linearities in the electron-beam energy profile may spoil the validity of the \textquotedblleft far-field\textquotedblright approach. However, as shown in Fig.~\ref{figure3b}, for a reasonable (i.e., close-to-experiment  \cite{fct}) electron-beam energy distribution, and using the already specified seed parameters, the agreement between the theoretical prediction and simulations is still very satisfactory.
\\
Figure~\ref{figure3c} shows the application of the proposed approach to the case in which the FEL pulse is split into two slightly saturated sub-pulses, as described in \cite{ninno}. Such a configuration has already been exploited at FERMI@Elettra for performing pump-probe experiments \cite{capotondi}. Remarkably, also in this case our prediction is able to reproduce the simulated pulse profile.
\\
Finally, Figure~\ref{figure3d} shows the effect of the envelope and the phase distortion induced by running the FEL in over-saturation regime. In this case, the value of the quadratic phase term is not large enough to make condition Eq.~(\ref{condition N}) valid, therefore the spectral profile does not reproduce the simulated temporal profile.
\begin{figure}[!h]
    \centering
    \subfigure
    {
        \includegraphics[clip=true, height=30mm]{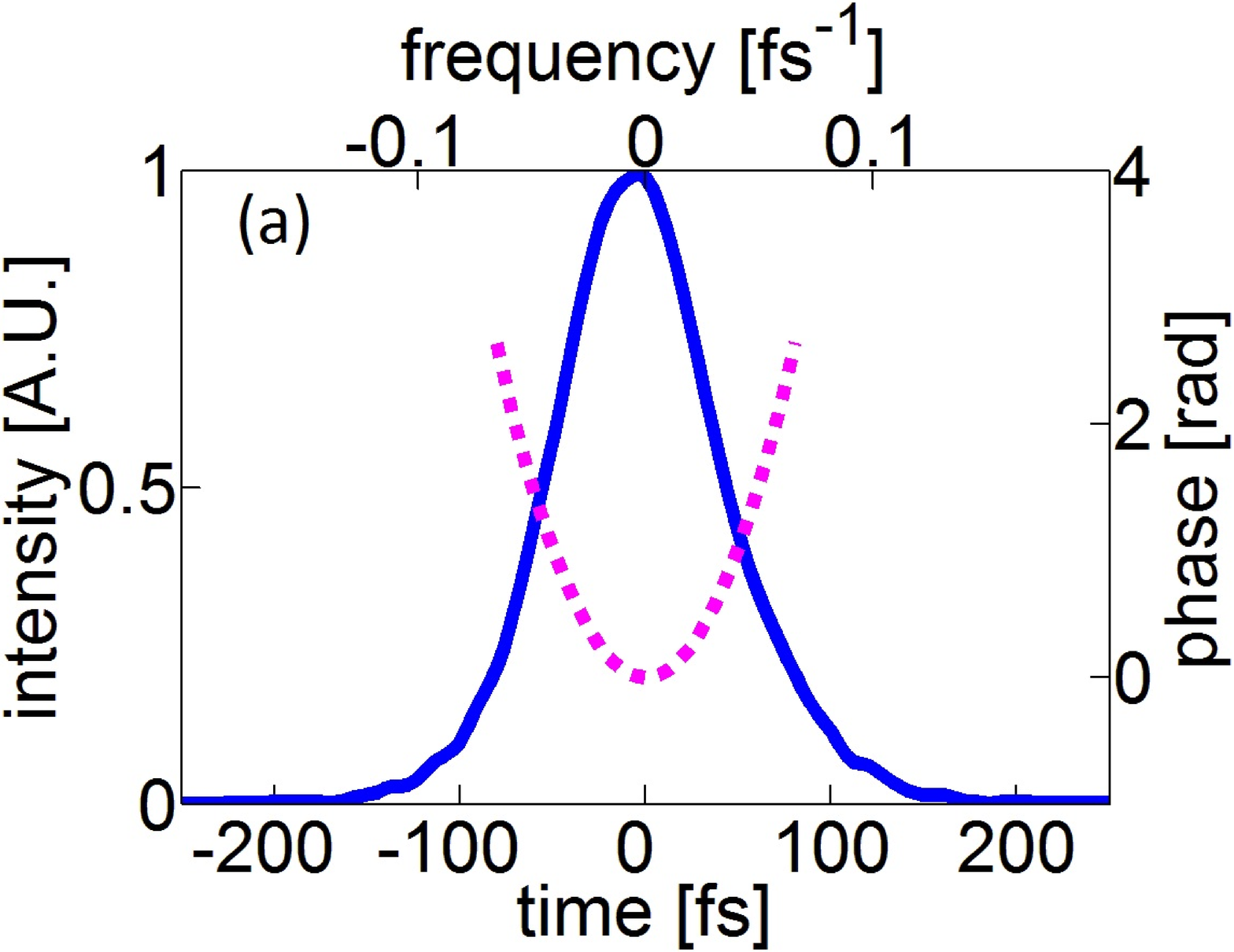}
        \label{figure4a}
    }
    \subfigure
    {
        \includegraphics[clip=true, height=25mm]{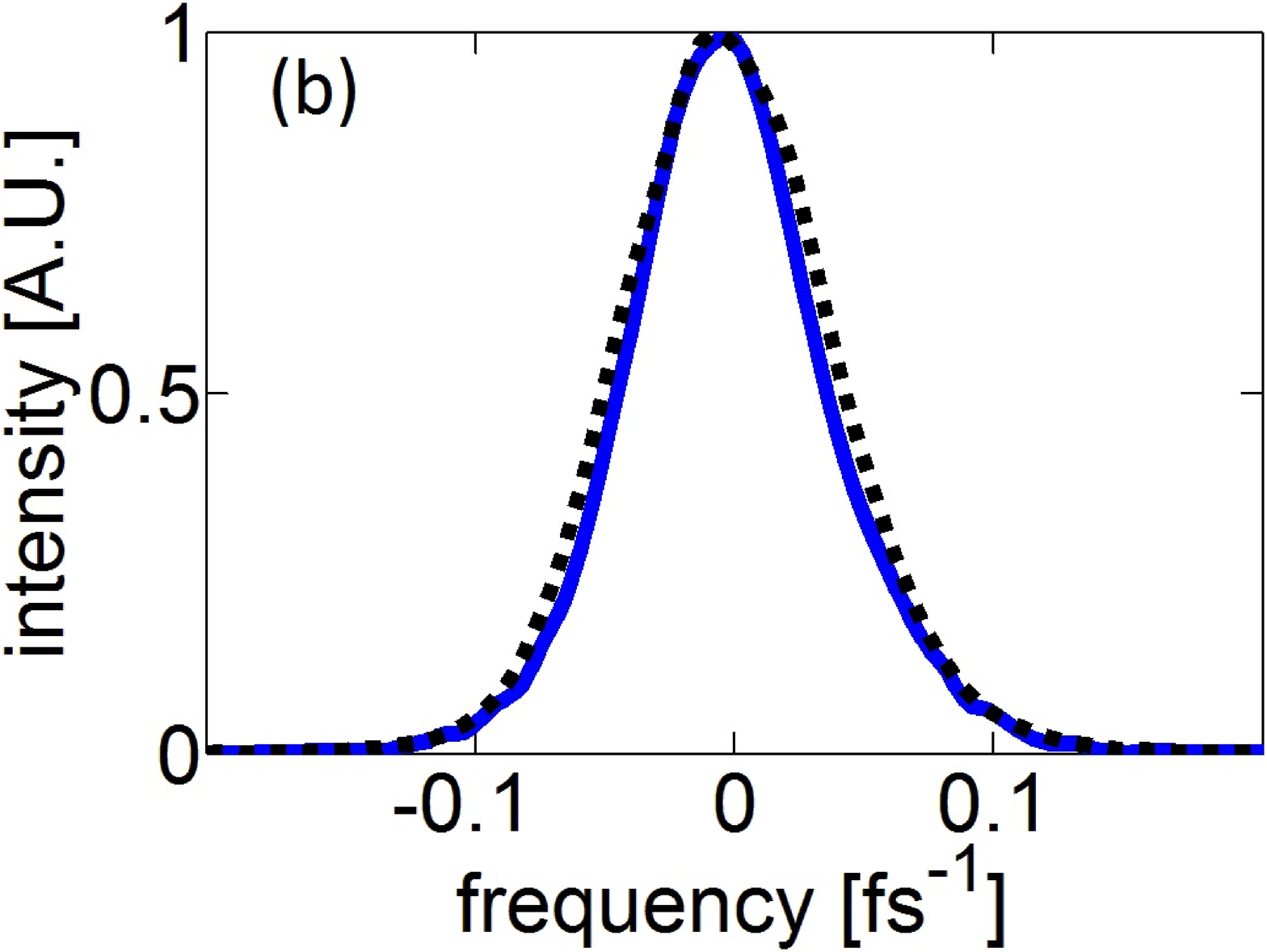}
        \label{figure4b}
    }
    \subfigure
    {
        \includegraphics[clip=true, height=30mm]{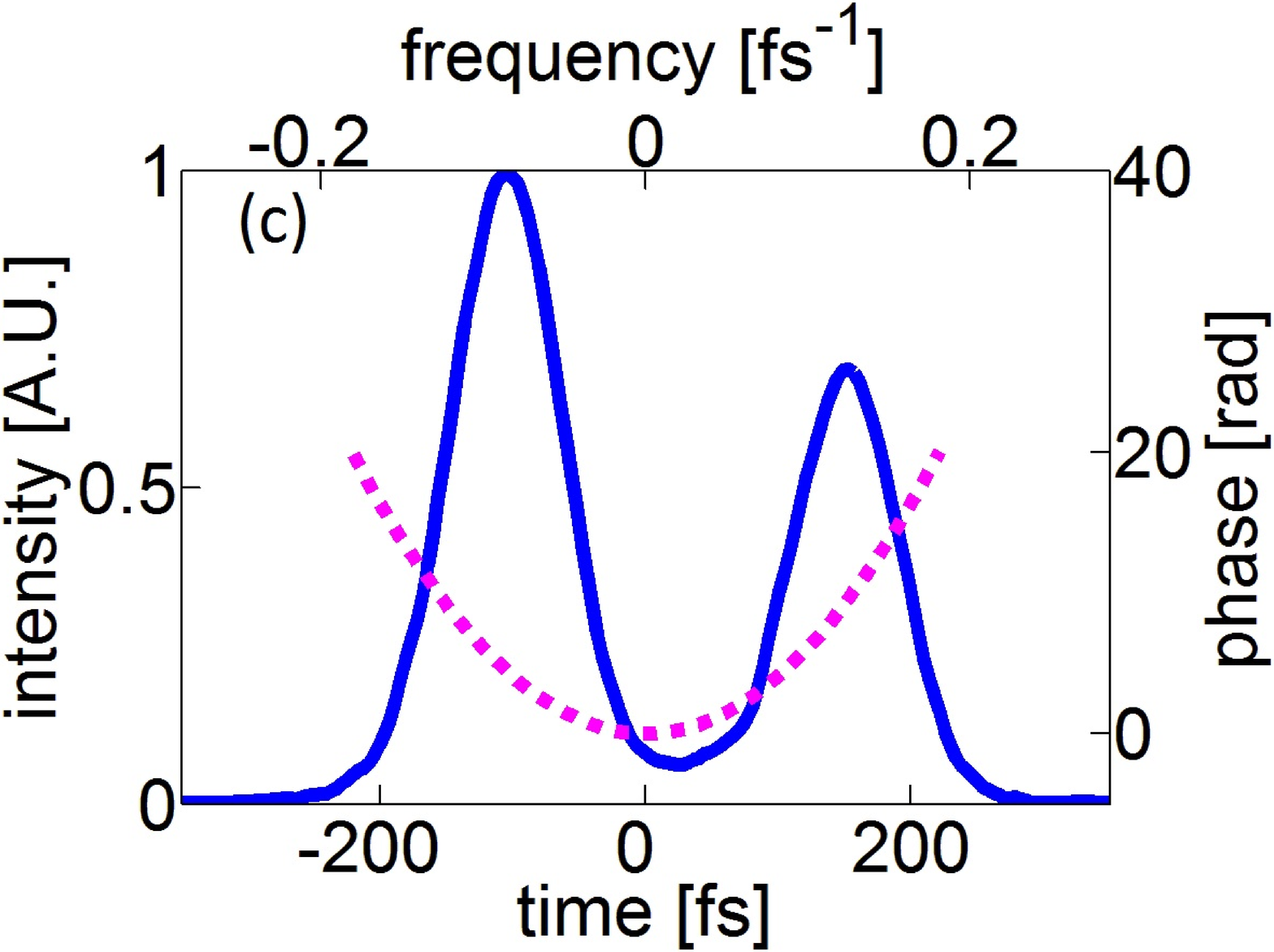}
        \label{figure4c}
    }
    \subfigure
    {
        \includegraphics[clip=true, height=25mm]{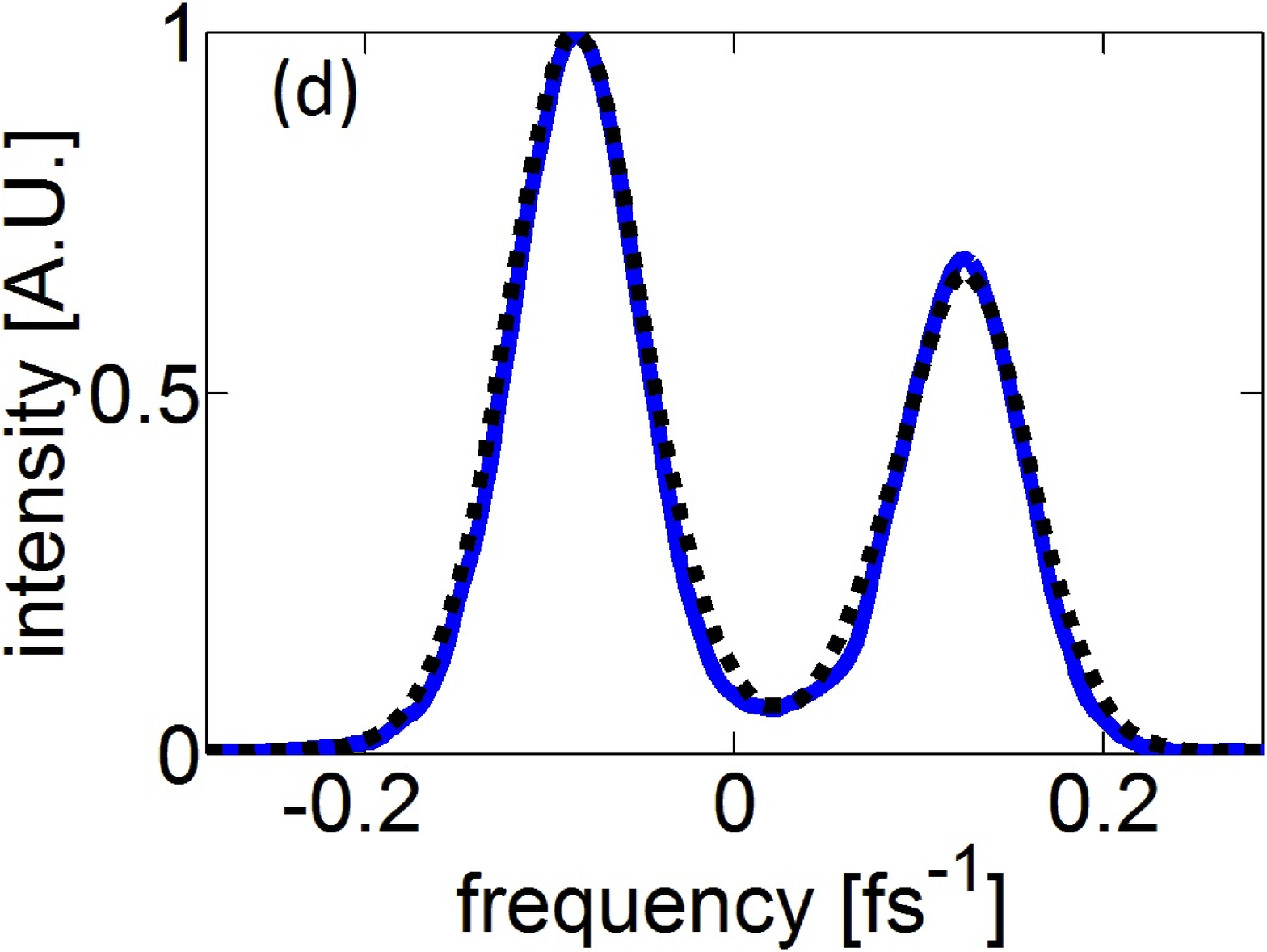}
        \label{figure4d}
    }
\FloatBarrier 
\caption{Retrieved temporal shape (solid blue line) from spectral measurement on FERMI@Elettra FEL: slightly saturated (a) and double-pulse (c) regimes. The temporal scale (bottom axis) is obtained from the spectral one (top axis), using the homothetic transformation given in Eq.~(\ref{homothetie}). The dot lines represent the quadratic phase term $n\Gamma_it^2$. Panels (b) and (d) show the comparison between the measured spectrum (solid blue line) and the reconstructed spectrum (dotted line), calculated using Eq.~(\ref{reconstruction}).}
\end{figure}
\\

Finally, we tested our approach by comparing theoretical results with spectral data from the FERMI@Elettra FEL source. The spectra were acquired by an on-line photon energy spectrometer \cite{svetina}, which uses the first-order of diffraction of a variable-line spacing grating allowing the beam focusing on a CCD detector. The main part of the beam (the zero order) goes unperturbed to the experimental station (see Fig.~\ref{fig2}).
\\
Figures~\ref{figure4a} and \ref{figure4c} display the temporal shape (bottom horizontal axis) retrieved from the measured spectra (top horizontal axis), using the homothetic relation Eq.~(\ref{homothetie}). For the considered case, the FEL was operated at 32 nm ($n=8$); as in the case of the simulation shown in Fig.~\ref{fig3}, the seed pulse had a measured (rms) pulse duration of $122\:\text{fs}$ and an intrinsic chirp $\Gamma_i=5.1 \:10^{-5}\:\text{fs}^{-2}$ (i.e., $N=0.16$). The measured value of the electron-beam energy curvature was $5\,\text{MeV/ps}^2$ (i.e., the same value used for the simulation shown in Fig.~\ref{figure3b}). It is worth noting that FERMI@Elettra has been operated under similar conditions also during user experiments. Figure \ref{figure4a} shows a pulse with quasi-Gaussian shape of 102 fs duration (FWHM), while Fig.~\ref{figure4c} refers to a case in which the FEL is operated in pulse-splitting regime.
\\
In order to corroborate our prediction, we compared the measured spectrum with the reconstructed one, $X_{REC}(\omega)$, defined as:
\begin{equation}
X_{REC}(\omega)=\left|\mathcal{F}_{\omega}\left\{\sqrt{X\left(2n\Gamma_i t\right)}e^{in\Gamma _i t^2}\right\}\right|^2
\label{reconstruction}.
\end{equation}
Here $\sqrt{X\left(2n\Gamma_i t\right)}$ is the retrieved complex amplitude, in which we suppose the hypothesis of negligible phase distortions. Obtained results, reported in Fig.~\ref{figure4b} and \ref{figure4d}, show a very satisfactory agreement. The reconstructed spectrum provides a validation on the retrieved pulse shape as well as on the absence of significant distortions and/or perturbations generated by any other noise source.
\\

In this letter, we established the possibility to determine directly the temporal profile of a frequency-chirped pulse from the measurement of its spectrum. This approach can be interested for any innovative light sources like the seeded FEL, as demonstrated here, or for example the so-called high harmonics generation (HHG). Light pulses produced by sources base on the harmonics generation of a driving laser carry generally an important frequency-chirp, even for a low chirp on the driving laser. Our approach exploits this intrinsic property, which may allows a non-invasive, shot-to-shot and real-time characterization of femtosecond XUV optical pulses.
\\

We thank the FERMI@Elettra commissioning team for providing the experimental data used in this paper. In particular, we acknowledge the contribution of the PADReS group headed by M. Zangrando and of the laser laboratory headed by M.B. Danailov. We also profited from insightful discussions with P. Rebernik Ribi\v{c}. This work has been partially supported by the project CITIUS, funded by the Cross-border Cooperation Programme Italy-Slovenia.
\\

\end{document}